# A Monte Carlo Study of the Relationship between the Time Structures of Prompt Gammas and in vivo Radiation Dose in Proton Therapy

**Wook-Geun Shin and Chul Hee Min\***

*Department of Radiation Convergence Engineering, Yonsei University, Wonju 220-710*

**Jae-Ik Shin, Jong Hwi Jeong, and Se Byeong Lee**

*Proton Therapy Center, National Cancer Center, Goyang 410-769*

For the in vivo range verification in proton therapy, it has been tried to measure the spatial distribution of the prompt gammas generated by the proton-induced interactions with the close relationship with the proton dose distribution. However, the high energy of the prompt gammas and background gammas are still problematic in measuring the distribution. In this study, we suggested a new method determining the in vivo range by utilizing the time structure of the prompt gammas formed with the rotation of a range modulation wheel (RMW) in the passive scattering proton therapy. To validate the Monte Carlo code simulating the proton beam nozzle, axial percent depth doses (PDDs) were compared with the measured PDDs with the varying beam range of 4.73-24.01 cm. And the relationship between the proton dose rate and the time structure of the prompt gammas was assessed and compared in the water phantom. The results of the PDD showed accurate agreement within the relative errors of 1.1% in the distal range and 2.9% in the modulation width. Average dose difference in the modulation was assessed as less than 1.3% by comparing with the measurement. The time structure of prompt gammas was well-matched within 0.39 ms with the proton dose rate, and this could enable the accurate prediction of the in vivo range.

Email: chmin@yonsei.ac.kr

Fax: +82-33-760-2892

# I. INTRODUCTION

The proton beams show the characteristic dose distribution, called as Bragg peak, delivering most of its dose before protons come to rest. The Bragg peak enables the highly conformal dose to the target volume while sparing the critical organs. Monitoring the location of the Bragg peak and its fall off is very important for the safety of the patient and for the successful treatment. Determining the in vivo dose distribution could be achieved with secondary particles (positron emitters and prompt gammas) generated with the proton-induced nuclear interactions and emitted out of the patient. Positron emission tomography (PET) may show the proton beam track along the beam direction by reconstructing the measured annihilation photons from the positron emitters of $^{11}C$ and $^{15}O$, however, it suffered from the low detection efficiency and biological wash-out [1]. It is reported that the prompt gammas isotropically emitted within $10^{-9}$ sec with the energy up to 10 MeV from $^{12}C^*$ and $^{16}O^*$ have the clear relationship with the proton dose distribution with much higher yield than positron emitters [2,3]. The prompt gammas could be measured during the beam on, and so the real time monitoring of patient dose is available in proton therapy. However, the proton beams passing through the beam nozzle including the range modulation wheel and compensator could generate the huge background radiations such as neutrons and photons, and this disturbs measuring the prompt gamma distribution.

There were many trials to measure spatial distribution of the prompt gammas with position sensitive detector such as array-type detector or Compton camera [4,5]. However, the spatial resolution showed limited accuracy and detection efficiency due to the high energy of the prompt gammas and the background radiations. The feasibility to resolve these problems was suggested with the collimation system and the time structure of the prompt gammas [6,7]. In passive scattering proton therapy, the range modulation wheel (RMW) rotates with 10 Hz of the rotating cycle and has the varied thickness on each step. The spread-out Bragg peak (SOBP) could be formed with the pristine beams passing through the RMW steps of the different thickness. According to the cycle of RMW, each position in the patient could have the characteristic time structure of the proton dose and the prompt gammas. In other word, characteristic time patterns of the prompt gammas measured in specific location could determine the residual range and the distal dose edge of the proton beams in the patient. Utilizing the time information of the prompt gammas allows the simple detection configuration for the point measurement with the sufficient collimation, which is limited to the position sensitive detection system.

The purpose of the current study is to evaluate the feasibility of the in-vivo and real time range verification technique using the time structure of the prompt gammas with RMW in passive scattering proton radiotherapy. This study employed the Proteus 235 proton beam nozzle code developed in the National Cancer Center (NCC) based on the Geometry and Tracking4 (Geant4) tool kit using the Monte Carlo (MC) method [8-10] and MC commissioning was performed by comparing the measured and simulated proton dose distribution in the water phantom.

## II. MATERIALS AND METHODS

**1. Modeling of Proton Beam Nozzle and Monte Carlo Commissioning**

For the validation of the Proteus 235 proton beam nozzle code, the axial percent depth doses (PDDs) of several beam conditions were calculated with different versions of the Geant4 tool kit in the water phantom of 50 x 50 x 50 cm$^3$ as shown in Fig. 1. The center line of the water phantom was divided into 1 mm voxel along the beam direction, and the entire deposited energy was calculated in each voxel. And the beam conditions were classified with its RMW track like B1 to B8 restricting the proton beam range and modulation. The ranges and modulation widths used in this study were shown in Table 1.

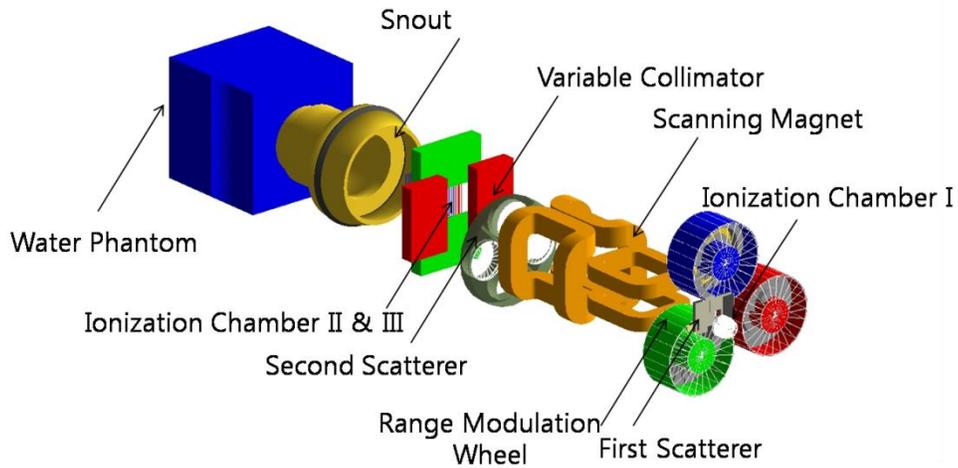

Fig. 1. The schematic illustration of the Proteus 235 proton beam nozzle and the water phantom simulated with the Geant4 tool kits.

Table 1. Proton beam ranges and modulations according to options indicating the RMW track

|    | Range (cm) | Modulation Width (cm) |
|----|------------|----------------------|
| B1 | 4.73       | 2.15                 |
| B2 | 6.11       | 2.78                 |
| B3 | 7.68       | 3.49                 |
| B4 | 9.74       | 4.43                 |
| B5 | 12.56      | 5.71                 |
| B6 | 16.75      | 7.61                 |
| B7 | 21.07      | 9.58                 |
| B8 | 24.01      | 10.21                |

The beam conditions and modulation options in the MC code were determined with the Conversion Algorithm (Convalgo) of the Ion Beam Applications (IBA) to simulate the same condition with the measurement using the beam nozzle system. Narrow proton beam was employed with10 cm of the field radius and 300 cm of source to surface distance (SSD) except B8. Note that B8 option was provided only when field radius is below 7 cm. The initial angle of the RMW was manually calibrated with the trial and errors, because the Convalgo does not currently support that condition. The tendency analysis was employed to find the correct initial angle of the RMW with Matrix Laboratory (MATLAB). The physics list of G4EMStandardPhysics_option3 for Electromagnetic (EM) process and QGSP_BIC_HP for hadron physics in the Geant4 code were used in the simulation [11]. The simulated PDD using the Geant4 version of 9.6.p02 and10.0 were compared with the PDD measured by the Markus ionization chamber.

## 2. Assessment of the Time Structure

The time structures of the proton dose and prompt gammas generated by the rotation of a RMW were assessed with the narrow proton beam of 180.28 MeV energy, 150 mm range, and 150 mm modulation width in the water phantom, and they were compared each other according to the depth in the water phantom. To simplify the simulation, it is assumed that the primary protons are delivered in the same position on the RMW with the same time and the RMW are discretely rotate with the angle of each step with the uniform speed of 100 ms. To calculate the time structure of the prompt gammas in the water phantom, several filters using G4Step and G4TrackVector class were used in storing information of the secondary particles as shown in Fig. 2.

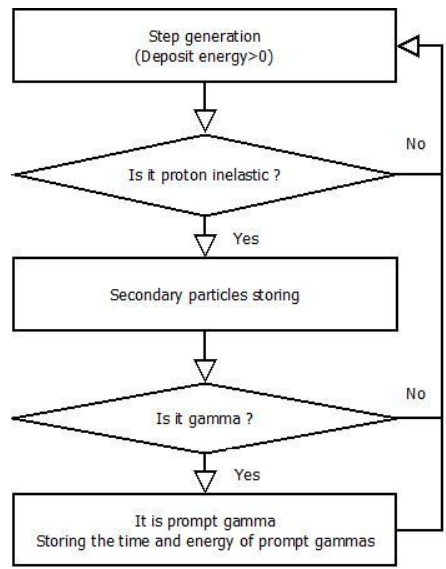

Fig. 2. Schematic diagram of the prompt gamma scoring algorithm by discriminating the other secondary radiation.

The G4Step class in the Geant4 code scored the total deposited energy and the physical process, and the G4TrackVector class chose the secondary particles generated by the proton-induced inelastic nuclear interaction. For the verification of the prompt gamma discrimination algorithm, the energy spectrum was compared with the theoretical gamma peak that could be observed in the water phantom.

### 3. Variance Reduction Technique

For the reduction of the simulation time, the phase-space file was employed. The primary proton beams were delivered to the water phantom, and then the particles passing through the whole modules in the nozzle was recorded in front of snout with the phase-space file which has information of the particles including the positions, directions, energies, type of particles and time information. In generating the phase-space file, only protons and neutrons which could influence to the prompt gamma distribution was considered. The number of the primary protons was $10^8$ and $3.1 \times 10^7$ particles were recorded in the phase-space file and the phase-space file was reused 10 times for the second run to calculate the time profile of the proton dose and prompt gammas in the water phantom. To reduce the calculation time, the energy cut-off was used for the electron transport, and the supercomputer called TACHYON in the Korea Institute of Science and Technology Information (KISTI) was also used [12]. The calculation time for the first run in generating the phase-space file was about 90 hour-processors and the second run in calculating the dose distribution was about 10 hour-processors.

### III. RESULTS AND DISCUSSIONS

**1. Monte Carlo Commissioning of Proteus 235 proton beam nozzle**

The simulated beam range and modulation width were compared with the measured data based on $D_{90}$ which means the location of the 90% of maximum dose. And the uniformity in the modulation width was directly compared in each depth. Without modifying the initial angle of the RMW, unacceptable difference between the measured and simulated dose distribution was found in the dose distal region. With the trials and errors, the initial angle of the RMW was calibrated by changing the angle of 4.12-13.00 degrees according to the beam conditions determining the ranges and modulations. With the initial angle calibration, the current results showed that the range difference and the average dose difference within the modulation region between the simulated and measured PDD for the ranges of 4.73-24.01 cm were assessed as less than 2.9% and 1.3%, respectively. Figure 3 shows the dose distribution of 8 cases out of 24 cases over the B1-B8 options for the measurement, Geant4.9.6, and Geant4.10. The PDD with different version of the Geant4 code was assessed as identical. However, the B7 option showed peak on 145 mm depth in the water phantom with maximum 3.37% difference. It is possible that the errors potentially come from the fluctuation of the beam current modulation (BCM), the inaccurate size of the initial proton beam, and the errors of the high energy cross-section in Geant4.

**2. Time Structure of Proton Dose and Prompt Gammas**

The energy spectrum of the prompt gammas was calculated in the water phantom to validate the discrimination algorithm. Figure 4 showed the simulated peaks of the prompt gammas caused by proton-oxygen inelastic nuclear interaction such as $^{16}O(p,x)^{12}C^{*4.439}$ and $^{16}O(p,x)^{15}N^{*5.299}$. Most of peaks were matched with the theoretical energy of the prompt gammas.

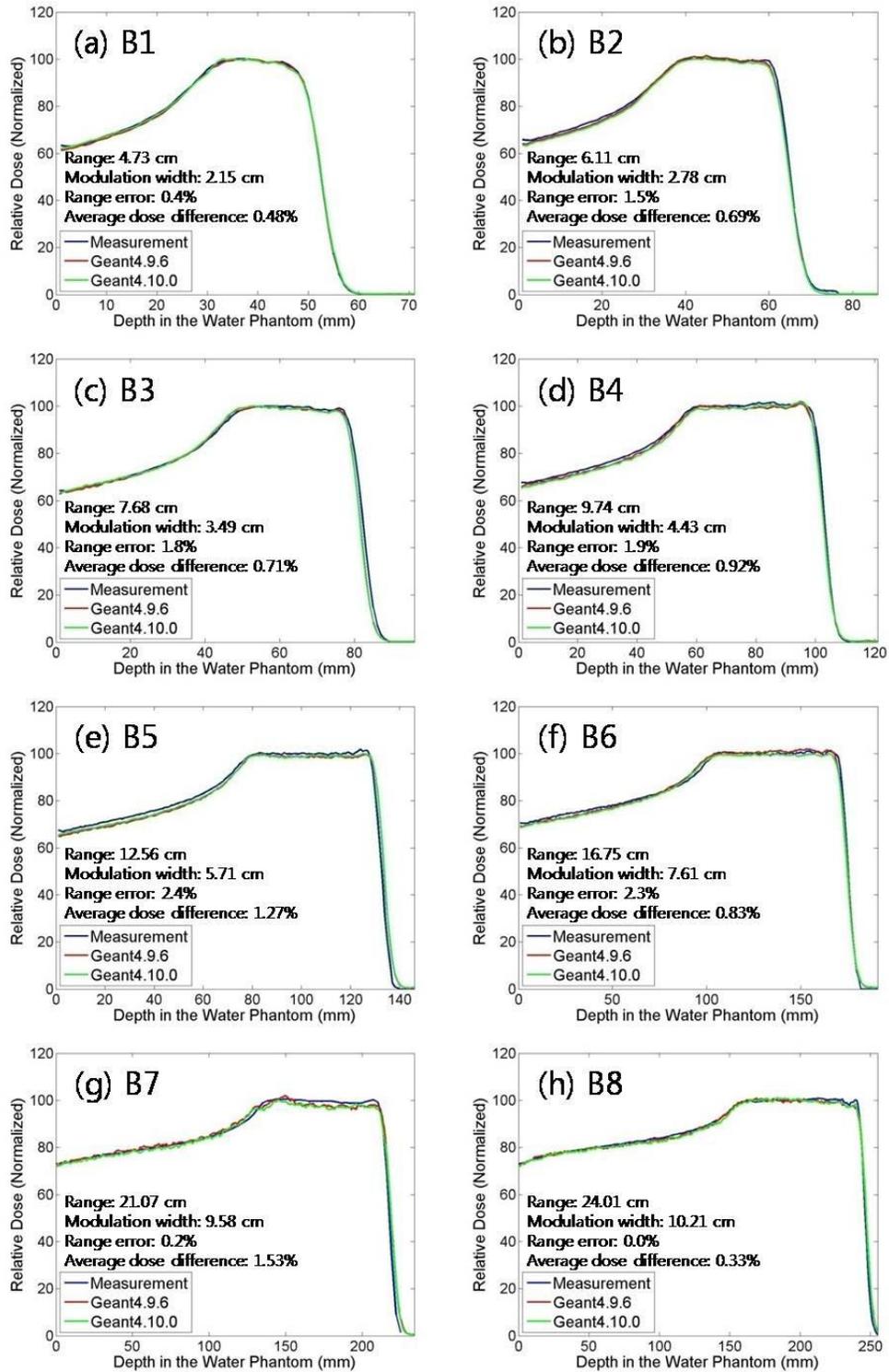

Fig. 3. Comparison of the measured PDD (green line) and simulated PDD with the Geant4.9.6 (red line) and Geant4.10.00 (blue line). The range and modulation options determined by B1 to B8 were shown with (a) to (h), respectively.

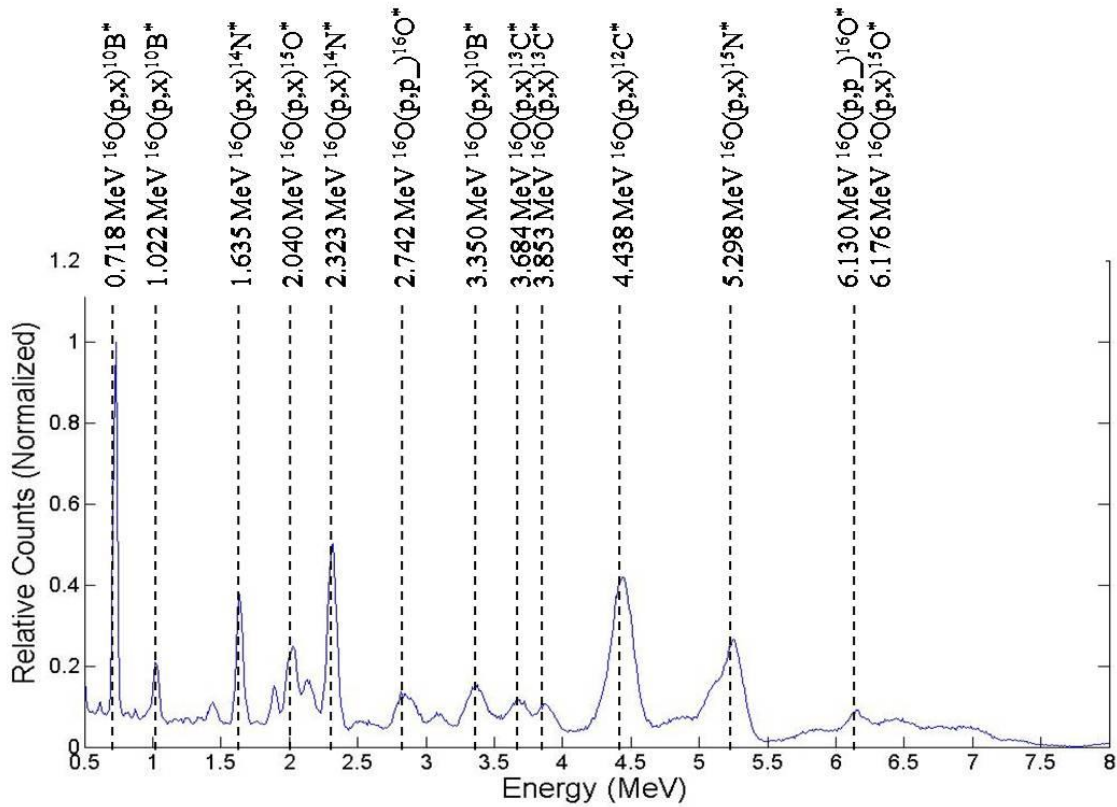

Fig. 4. The energy spectrum of prompt gammas calculated with the Geant4 and the theoretical peak according to the interaction types.

With the algorithm, the time structure of the prompt gammas and the proton dose were assessed at each depth of the water phantom. Nevertheless, the time resolution was 0.39 ms with the discrete rotation of the 256 steps with the RMW frequency of 100 ms, the relative fall-off difference of the time structure were assessed as less than onetime-bin (0.39 ms) as shown in Fig. 5. The optimal scale in the prompt gamma distribution could indicate the well-matched fall off with the proton dose distribution. Note that, Figure 5 shows the reduced proton dose to 25% of its original values just for the visualization purpose lather than increasing the prompt gamma values. Theoretically, the proton-induced nuclear interactions show the threshold energy of about 15 MeV in generating the prompt gammas, and this make the slight difference between the prompt gamma and proton dose distributions.

However, reducing the proton dose to 25% of its original values showed the more clear relationship at the end of the curves in the Fig. 5.

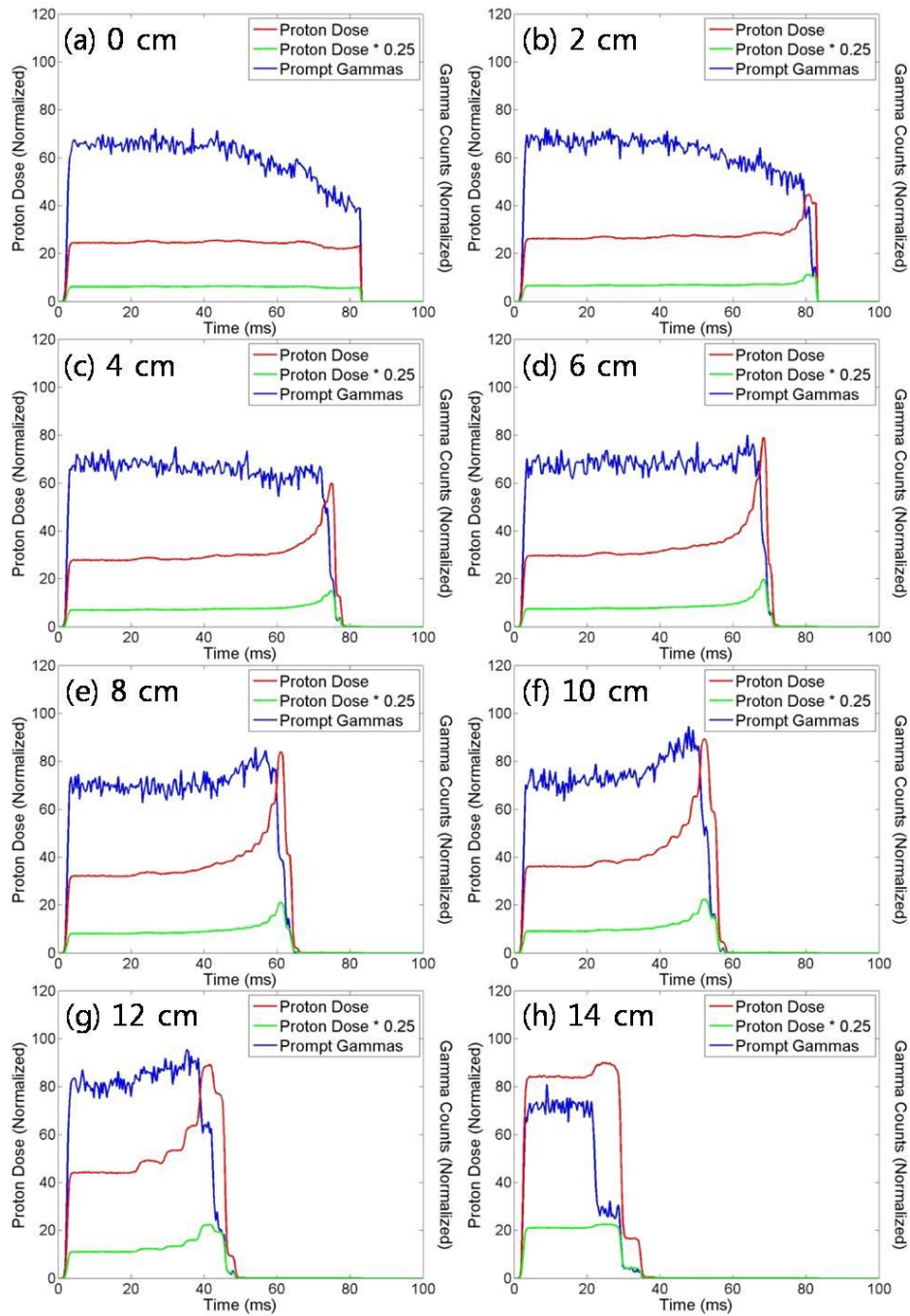

Fig. 5. The time structure of proton dose (red line), proton dose x 0.25 (green line), and prompt gammas (blue line), in 0 cm (a), 2 cm (b), 4 cm (c), 6 cm (d), 8 cm (e), 10 cm (f), 12 cm (g), and 14 cm (h) depth of the water phantom.

## IV. CONCLUSION

The current study demonstrated the feasibility of the range verification technique by measuring the time structure of the prompt gammas in the passive scattering proton therapy with the Monte Carlo method using the Geant4 tool kit. It is expected that the time information in measuring the prompt gammas may clearly reduce the background gammas with the time window in the experiment. The in vivo validation technique utilizing the time structure could be applied to not only the passive scattering but also the active scanning method in proton therapy. The proton spots in the active scanning also should have the characteristic time structure according to the depth by proton beam scanning patterns. The MC commissioning also validate the identical results of the different version of the Geant4 and the proper modeling in implementing the proton beam nozzle. The developed beam nozzle code using Geant4 could be employed to the other various researches of the proton therapy such as proton radiography, RTP validation, and biological effect of the proton beams. However, the mismatch in the B7 option is still remaining limitation. Based on the current results, the optimization study for the prompt gamma detector would be followed to configure the prototype detection system. Increasing the detection efficiency and decreasing the background gammas are critical factors in the optimization study.

## ACKNOWLEDGEMENT

This research was supported by Basic Science Research Program through the National Research Foundation of Korea (NRF) funded by the Ministry of Science, ICT & Future Planning (NRF-2014R1A1A1007789) and Korea Institute of Nuclear Safety funded by the Nuclear Safety and Security Commission, Republic of Korea.